\documentstyle[camera,jjapplug,psfig]{article}
\makeatletter

\catcode`\_\active
\let\@orgsb\sb
\def\@newsb#1{\ifmmode \@orgsb{#1}\else $\@orgsb{#1}$\fi}
\def_{\protect\@newsb}

\catcode`\^\active
\let\@orgsp\sp
\def\@newsp#1{\ifmmode \@orgsp{#1}\else $\@orgsp{#1}$\fi}
\def^{\protect\@newsp}

\makeatother

\title{An Improved Method for Obtaining Single-Phase Sr_2MoO_4 under Controlled Ultralow Oxygen Partial Pressure}

\author{Naoki~{\sc Shirakawa}\thanks{E-mail address:~shirakawa.n@aist.go.jp}, Shin-Ichi~{\sc Ikeda}, Hirofumi~{\sc Matsuhata} and Hiroshi~{\sc Bando}}
\inst{Nanoelectronics Research Institute, AIST, Tsukuba 305-8568, Japan}

\recdate{April 25, 2001}
\revdate{June 4, 2001}
\accdate{June 6, 2001}

\abst{%
We investigated 
a method for obtaining a sintered bulk sample of single-phase Sr_2MoO_4, which has recently increased in 
importance due to its similarities to
superconducting Sr_2RuO_4. We discovered that with the aid of Ti_2O_3 as a
$p({\rm O}_2)$ buffer, high-purity Sr_2MoO_4 is successfully synthesized in a sealed tube. 
We confirmed the space group and determined the exact elemental composition of the compound in order to establish 
its stoichiometry.
The guideline for the rational choice of a $p({\rm O}_2)$-buffer material is discussed.
}
\kword{Sr_2MoO_4, layered molybdate, Sr_2RuO_4, synthesis, K_2NiF_4 structure, $p({\rm O}_2)$ buffer, Ti_2O_3}

\begin{document}
\maketitle


For many years there had been 
debates over the existence of the compound Sr$_2$MoO$_4$\cite{scholder,balz,mccarthy} until Lindblom and Ros\'en described their method
for synthesizing it and 
reported that it crystallizes in a K$_2$NiF$_4$-type structure
\cite{lindblom,jcpds}.
What they actually obtained, however, was the mixture of Sr$_2$MoO$_4$ and SrO, and the
latter 
had to be washed away with methanol. 
Recently, renewed interest in Sr$_2$MoO$_4$ has emerged in 
 solid-state physics 
  with 
the discovery
of unconventional superconductivity in Sr$_2$RuO$_4$\cite{maenoNature,maenoReview}.
The need for a bulk sintered sample 
rather than a
powder sample in many physical measurements urged 
 the development of  sintered  
 Sr$_2$MoO$_4$ bulk, which is single-phased as-prepared. 
Moreover, the paper of Lindblom and Ros\'en does not provide the detailed information necessary for a cross-examination.
Recently, Steiner and Reichelt attempted 
to 
reproduce 
their result\cite{steiner}, but were only able to obtain samples containing relatively large amounts of Mo metal and other phases.  We will consider the reason later. This large amount of Mo metal makes it difficult to look into 
the possibility of superconductivity in Sr$_2$MoO$_4$ below 0.9\,K\@, the $T_{\rm c}$ of Mo metal.

Here 
we describe 
our new method that 
solves all of these problems and enables 
us 
to obtain a single-phase sintered bulk of
Sr$_2$MoO$_4$ with ease and perfect 
reproducibility. 
In addition, 
this procedure 
gives 
new 
life 
to the conventional sealed-tube method and has the potential for 
creating
many new materials in the future.

\begin{figure}
\centerline{\psfig{figure=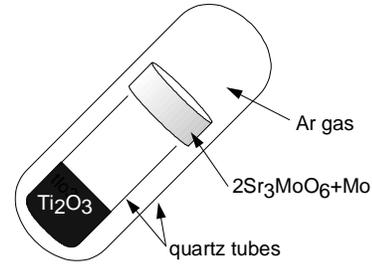,height=3.5cm}}
\caption{Schematic 
of the sealed tube for reaction.}
\label{f.quartztube}
\end{figure}

As illustrated in Fig.~\ref{f.quartztube}, we placed a pellet of the $2:1$ molar mixture of
Sr_3MoO_6 and Mo metal powder (99.9\%) on top of an open inner quartz tube holding Ti_2O_3 powder inside. An outer quartz tube 
with the above contents was first evacuated and
then charged 
with Ar gas at a 
pressure that was going to be 1\,atm at the firing
temperature. Finally, the outer tube was sealed by fusion. The optimal molar amount of Ti_2O_3
turned out to be 
0.75 per mole of Sr_2MoO_4 after several trials. The sealed tube was kept at 1150\,$^\circ$C for one 
week. Prior to this, 
Sr_3MoO_6 had been prepared
according to the reaction $\rm 3SrCO_3+MoO_3 \rightarrow Sr_3MoO_6+3CO_2\uparrow$. The mixture of SrCO_3 (99.99\%) and MoO_3 (99.99\%) was 
heated in flowing Ar atmosphere at 1000\,$^\circ$C for 24\,h with two intermittent grindings.

Samples were examined 
for phase purity by powder X-ray diffraction (PXRD) at room temperature with Cu-$K\alpha$ radiation.  Electron diffraction (ED) was performed to determine the space group.
The lattice constants were calculated from the peak positions in the PXRD pattern and the Miller indices that had been assigned based on ED results.
The ratio of metal elements (Sr to Mo) was measured by an inductively coupled plasma atomic
emission spectrometer (ICP-AES). The oxygen content was deduced from the result of thermogravimetric analysis (TGA). The TGA was performed in an atmosphere of $4:1$-volume mixture
 of argon and oxygen.



\begin{figure}
\centerline{\psfig{figure=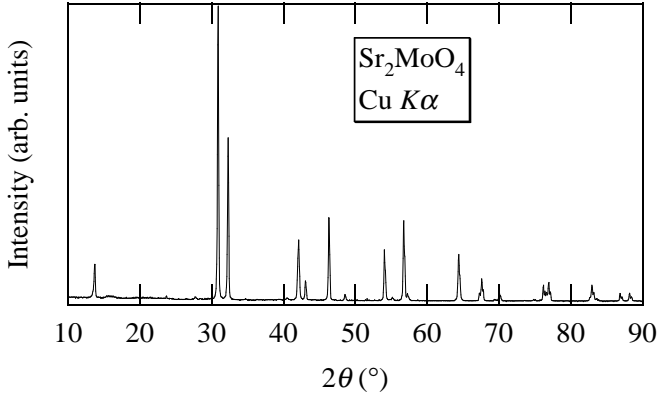,width=\columnwidth}}
\caption{PXRD 
pattern of a Sr_2MoO_4 sample. All peaks other than 
the tiny 
one at 40.5\,$^\circ$ belong to Sr_2MoO_4.}
\label{f.xrd}
\end{figure}

We show a 
PXRD pattern of a typical Sr_2MoO_4 sample, obtained according to the above
procedure, in Fig.~\ref{f.xrd}.
The only impurity phase visible in this figure is a 
 tiny peak at 40.5\,$^\circ$,
which we attribute 
to Mo metal. Basic physical properties, such as electrical
resistivity, magnetic susceptibility, and specific heat, of our samples have been reported elsewhere\cite{ikedaJPSJ,ikedaPhysicaC}. No superconductivity has been observed as yet even down to a temperature of
30\,mK\@.

\begin{figure}
\centerline{\psfig{figure=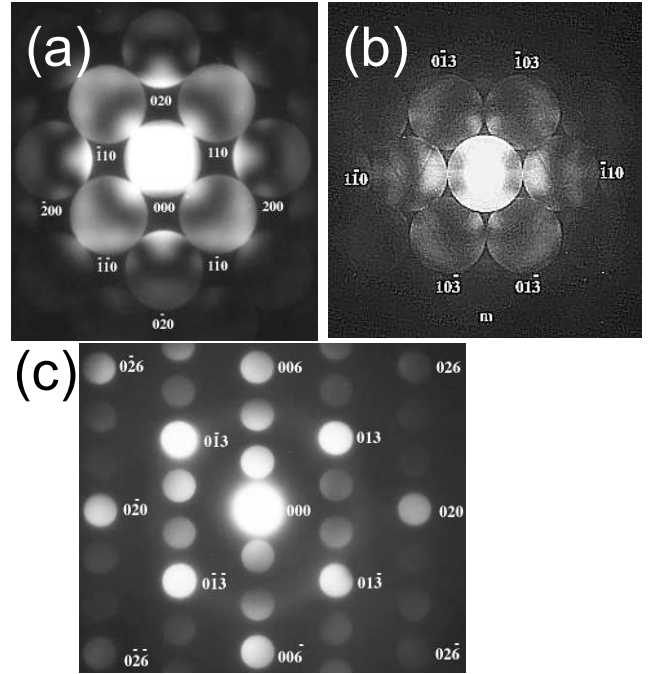,height=9cm}}
\caption{CBED 
patterns with the incident beam perpendicular to the
(a)~(001)-, 
 (b)~(113)-, 
  and (c)~(100)-plane. The mark ``m'' in (b) denotes
the mirror plane. The indices are assigned for the $I4/mmm$ symmetry.}
\label{f.ed}
\end{figure}

Figure~\ref{f.ed} shows three patterns of convergent-beam electron diffraction
(CBED) on our sample of Sr_2MoO_4. They
are all consistent with the space group $I4/mmm$, ensuring that its crystal
structure is actually tetragonal K_2NiF_4-type, as first suggested by Balz and
Plieth\cite{balz}. We indexed the peaks in the PXRD pattern 
and calculated the lattice parameters. Our results, $a=3.9168(4)$\,\AA\ and $c=12.859(2)$\,\AA, are more consistent with the lattice parameters of 
Steiner
and Reichelt\cite{steiner} than those of Lindblom and Ros\'en\cite{lindblom}.

\begin{figure}
\centerline{\psfig{figure=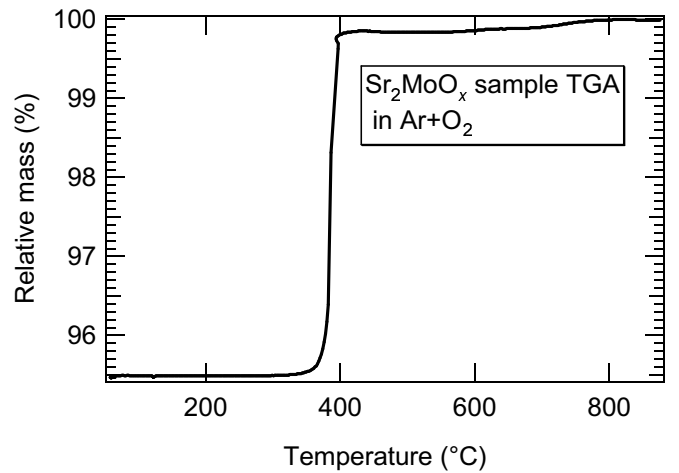,width=\columnwidth}}
\caption{The change of mass of a Sr_2MoO_x sample during heating in oxidizing atmosphere.}
\label{f.tga}
\end{figure}

We determined the ratio of $\rm Sr:Mo$ to be $2.02\pm0.01$ by an ICP-AES\@
. The result of TGA 
is shown in Fig.~\ref{f.tga}. Assuming that all the Mo^{4+} ions had become Mo^{6+} species by heating 
to 840\,$^\circ$C, where
 a plateau was achieved, 
the oxygen content $x$ in Sr_{2.02\pm0.01}MoO_x turned out to be $4.001\pm0.003$.
We may safely say that the composition of what has been called
Sr_2MoO_4 is actually $2:1:4$.
We also note that no appreciable oxygen nonstoichiometry exists in this system,
at least on the reduction side of Mo^{4+}, by comparing 
the peak positions in the PXRD 
patterns with various amounts of Mo metal precipitation.

Lindblom and Ros\'en have measured the oxidation potential for the 
equilibrium
$ \rm 2SrO(s) + Mo(s) + O_2 (g) 
 \;\vcenter{\hbox{$
\stackrel{\textstyle\rightarrow}{\textstyle\leftarrow}$}}\;
 Sr_2MoO_4(s)
$
and concluded that the equilibrium oxygen partial pressure $p({\rm O}_2)$ 
is $10^{-21}$\,atm at 1200\,K\cite{lindblom}.
To maintain 
this extremely low $p({\rm O}_2)$
, they used the mixture
of SrO and Mo metal as the $p({\rm O}_2)$ buffer, which was placed on top of 
the starting
materials in a sealed Al$_2$O$_3$ tube. However, they did not report 
the ratio of the amount of the buffer 
to that of the starting materials. Apart from this, their starting materials were the $3:1$ molar mixture of
$\rm Sr(OH)_2$ and MoO_2, which necessitated the separation of SrO from the
resultant mixture.

Steiner and Reichelt started from the stoichiometric $(2:1)$ mixture of SrO and MoO$_2$, which was sealed in a quartz tube along with a $p({\rm O}_2)$ buffer of SrO and Mo\cite{steiner}. It resulted in
a mixed-phase sample containing Sr$_2$MoO$_4$ (the main phase), Mo, SrMoO$_3$
and so forth. We can understand this in the following way: The starting materials ($\rm 2SrO+MoO_2$) and the $p({\rm O}_2)$ buffer ($\rm SrO+Mo$) can exchange $\rm O_2$
mutually. 
As a consequence of thermal equilibrium, metallic Mo or an oxide containing Mo^{3+} should 
 emerge 
 on the 
starting material's side. In other words there is no reason for 
all of the Mo ions in the starting materials to remain $4+$ in valence, while there is 
metallic Mo on the buffer's side and oxygen can be exchanged between both sides.

In Lindblom and Ros\'en's case, another equilibrium, $\rm 2H_2O\;
\vcenter{\hbox{$\stackrel{\textstyle\rightarrow}{\leftarrow}$}}\;2H_2+O_2$, was also involved thanks to the
usage of Sr(OH)_2. 
This may have helped 
maintain $p({\rm O}_2)$ at the level
suitable for Sr_2MoO_4 formation.  The condition seems, however, so subtle that
we could not 
obtain any Sr_2MoO_4 according to 
their procedure. 

We began investigating 
an alternative method for 
obtaining the target compound with more ease and reliability\cite{shirakawa}. 
Another goal was to produce phase-pure Sr_2MoO_4 samples that need no postproduction treatment for removing other phases.
Thus, 
we fixed the starting mixture at 
$\rm 2Sr_3MoO_6 + Mo$, which
was to yield 
3Sr_2MoO_4 only.
We concentrated on finding the right $p({\rm O}_2)$-buffer material using a logical and
rational approach.
We emphasize that sealing the starting mixture under high vacuum of $\sim10^{-5}$\,Torr and firing
it produced no Sr_2MoO_4, as had already been observed by McCarthy and Gooden%
\cite{mccarthy}.
Lower 
$p({\rm O}_2)$ than can be achieved 
by usual pumping was required
, as mentioned by Lindblom and Ros\'en\cite
{lindblom}.

To simply reduce $p({\rm O}_2)$, the use of an oxygen-getter element is effective.
We tried putting Ta foil in the sealed tube along with the starting mixture.
It resulted in a large amount of Mo metal, seemingly due to the too low $p({\rm O}_2)$
in the tube. Since Ta has a stronger ionization tendency than Mo does, Ta foil
may have absorbed not only residual oxygen in the tube, but also oxygen in the starting
materials and made Sr_2MoO_4 unstable.
Our objective then
 was to highly stabilize Mo^{4+} ions in order to promote
the formation of Sr_2MoO_4, which contains Mo as Mo^{4+} ions.

This is where
the concept of ionization tendency of {\it ions} comes in. It is an
extended version of ionization tendency of metallic elements. Let us present an
example. A Mo^{6+} ion cannot be ionized further by any chemical means.
On the other hand, elemental Mo is 
  ionized more easily than 
all the other Mo ions.
A Mo^{4+} ion falls between these two extremes. Consequently, we can
draw an ionization tendency of ions stating $\rm Mo^{6+}<Mo^{4+}<Mo^0$.

When choosing the buffer material to stabilize Mo^{4+}, the following two points must be taken into consideration. 1)~The buffer should be weaker than Mo^0 in the ionization
tendency. This ensures that the buffer does not reduce Mo^{4+} to Mo metal.
2)~The buffer should be stronger than Mo^{4+} in the ionization tendency. 
If this condition is met, the buffer will act as a weak 
reducing agent on the compound
containing Mo^{4+}, thereby suppressing the absorption of oxygen into the target compound and its decomposition due to oxidation.

What kind of buffer material can satisfy these requirements? We expanded 
the range of candidates from metallic elements to oxides and chose 
Ti_2O_3, containing Ti^{3+}. Judging from the tables for 
oxidation potentials of various ions,\cite{knacke} the condition $\rm Mo^{4+}<Ti^{3+}<Mo^0$ was apparently met. We decided to use Ti_2O_3
as the buffer material.
This was the main part of our technique that supplied a high-purity bulk of
Sr_2MoO_4, as manifested in Fig.~\ref{f.xrd}.

The role of Ar gas sealed in the quartz tube is not yet understood. However, without Ar gas, the sample did not become as pure as 
 in Fig.~\ref{f.xrd} under the same firing condition. We speculate
that Ar works as a carrier gas for oxygen molecules. Since $p({\rm O}_2)$ in the tube is extremely low, an oxygen molecule is rarely 
scattered by other oxygen molecules. If the total pressure
in the tube is also low, oxygen molecules travel 
linearly over 
 a long distance and
collide infrequently with both the starting mixture and the buffer. 
In this situation it will take very long for the whole system to reach equilibrium. Argon gas may contribute to increasing the efficiency of the exchange of oxygen between the starting materials
and the buffer.

Actually we are not the first to use a transition-metal oxide in order to make
a complex oxide of another transition metal. Nozaki {\it et al.}\cite{nozaki}\ used TiO as a ``reducing agent'' for making Sr_2VO_4. They switched the agent to Ti_2O_3 for preparing Sr_3V_2O_7. This paper of ours, however, explains how and why their
method worked and gives the guidelines for rationally choosing a $p({\rm O}_2)$-buffer material for synthesizing other difficult-to-obtain oxides.

In conclusion, we have succeeded in synthesizing truly phase-pure Sr_2MoO_4,
which is useful for physical-properties measurements. It was necessary to control $p({\rm O}_2)$ in the sealed tube at 
an extremely low level that stabilized Sr_2MoO_4 and we solved this problem by using Ti_2O_3
as the $p({\rm O}_2)$ buffer. We described how to choose a buffer material rationally,
which was based on the concept of the ionization tendency of {\it ions}. 
The oxidation potential data of various ions have been  useful.

We thank I. Hase for valuable discussion.

\makefigurecaptions


\begin{thebibliography}{12}
  \bibitem{scholder} R.~Scholder: Angew. Chem. {\bf66} (1954) 461.
  \bibitem{balz} D.~Balz and K.~Plieth: Z. Elektrochem. {\bf59} (1955) 545.
  \bibitem{mccarthy} G.~J.~McCarthy and C.~E.~Gooden: J. Inorg. Nucl. Chem. {\bf35} (1973) 2669.
  \bibitem{lindblom} 
  B.~Lindblom and E.~Ros\'en: Acta Chem. Scand. A {\bf40} (1986) 452.
  \bibitem{jcpds}
 JCPDS card 45-80.
  \bibitem{maenoNature} Y.~Maeno, H.~Hashimoto, K.~Yoshida, S.~Nishizaki, T.~Fujita, J.~G.~Bednorz and F.~Lichtenberg: Nature {\bf372} (1994) 532.
  
  \bibitem{maenoReview} 
Y.~Maeno, T. M.~Rice and M.~Sigrist:
Phys. Today {\bf54} (2001) 42 and references therein.

  \bibitem{steiner} U.~Steiner and W.~Reichelt: Z. Naturforsch. {\bf53b} (1998) 110.
  
  \bibitem{ikedaJPSJ} S.~I.~Ikeda, N.~Shirakawa, H.~Bando and Y.~Ootuka: J.
  Phys. Soc. Jpn. {\bf69} (2000) 3162.
  
  \bibitem{ikedaPhysicaC} S.~I.~Ikeda and N.~Shirakawa:
  Physica C {\bf341-348} (2000) 785.
  \bibitem{shirakawa} N.~Shirakawa and S.~I.~Ikeda: Physica C {\bf341-348} (2000) 783.
  
  \bibitem{knacke} Tables of oxidation potential of selected elements and
ions are found in many handbooks and textbooks. One of the most exhaustive 
lists is {\it Thermochemical Properties of Inorganic Substances}, eds. O.~Knacke,
O.~Kubaschewski and K.~Hesselmann (Springer-Verlag, Berlin, 1991) 2nd ed.

  \bibitem{nozaki} A.~Nozaki, H.~Yoshikawa, T.~Wada, H.~Yamauchi and S.~Tanaka:
  Phys. Rev. B {\bf43} (1991) 181.
\end{thebibliography}
\end{document}